\documentclass[12pt]{article}
\def \beq {\begin{equation}}
\def \eeq {\end{equation}}
\def \beqn {\begin{eqnarray}}
\def \eeqn {\end{eqnarray}}
\begin{document}

\centerline{\bf \Large{Vortical susceptibility of finite-density QCD matter}} 
\vspace{1cm}
\centerline{\bf A. Aristova$^1$, D. Frenklakh$^{1,3}$, A. Gorsky$^{2,3}$, D. Kharzeev$^{4,5}$ }
\vspace{1cm}
\begin{centerline}
{\it $^{1}$ Institute of Theoretical and Exprerimental Physics, Moscow, Russia}

{\it $^{2}$ Institute of Information Transmission Problems of the Russian Academy of Sciences,
Moscow, Russia,}

{\it $^{3}$ Moscow Institute of Physics and Technology, Dolgoprudny , Russia}

{\it $^4$Department of Physics and Astronomy, Stony Brook University, NY 11794-3800, USA}

\noindent{\it $^5$Physics Department and RIKEN-BNL Research Center, \\ Brookhaven National Laboratory, NY 11973-5000, USA}
\end{centerline}
\vspace{1cm}

\centerline{\bf \large{Abstract}}
The susceptibility of finite-density QCD matter to vorticity is introduced, as an analog of magnetic susceptibility. It describes the spin polarization of quarks and antiquarks in finite-density QCD matter induced by rotation. We  estimate this quantity in the chirally broken phase using the 
mixed gauge-gravity anomaly at finite baryon density. 
It is proposed that the vortical susceptibility of QCD matter is responsible for the polarization of 
$\Lambda$ and  ${\bar \Lambda}$ hyperons observed recently in heavy ion collisions at RHIC by the STAR Collaboration.

\section{Introduction}

Finite-density relativistic matter exhibits a rich variety of macroscopic phenomena that stem from quantum 
anomalies involving the baryon and electric charge densities \cite{zhit1,zhit2,zhit3}. 
These anomalies lead in particular to the Chiral Vortical Effect (CVE) identified both in field theory \cite{zhit3} and in the holographic approach \cite{erdmenger}; the temperature dependence of the effect is non-trivial and originates from the mixed gauge-gravity anomaly \cite{cve}. The anomaly-induced effects of rotation are analogous to effects in an external magnetic field, albeit with some important differences \cite{Kharzeev:2010gr}.

In this paper we will extend the previous analyses to the effect of rotation on finite-density QCD matter in the phase with spontaneously broken chiral symmetry and the corresponding gap in the excitation spectrum.
The response of the ground state of a gapped system 
to external fields is an important property of the system. A familiar
example is magnetic susceptibility of quark condensate in QCD
introduced in \cite{ioffe} that measures the linear response of vacuum
to the external magnetic field. Its anomalous nature has been recognized in \cite{vainshtein}
and it has been evaluated in holographic QCD taking into account the 
conventional Chern-Simons (CS) term \cite{gk} or the mixed CS term in the extended holographic model 
\cite{gkkv,harvey}.

The magnetic susceptibility of QCD in the confined  phase has been evaluated 
on the lattice \cite{buividovich,bali} at finite temperature. It has also been shown 
that magnetic susceptibility is responsible for a number of interesting physical phenomena, including 
the T-odd correlations in dijet production \cite{braun} and the radiative decays of heavy mesons \cite{rohrwild}.

A number of physical systems possess vorticity -- for example, the quark-gluon plasma produced in heavy ion collisions at a non-zero impact parameter, or dense matter in rotating neutron stars.
A natural question arises: is there an analog of the magnetic susceptibility of QCD matter describing the response of the system to rotation?

In this paper we demonstrate that there is indeed a new anomalous effect on QCD matter in spontaneously broken phase that is induced by rotation; it can be describes as a  nontrivial 
response of the quark condensate to the external gravitational field. The simplest
non-flat gravity background involves the abelian graviphoton gauge connection. The 
corresponding gravimagnetic curvature corresponds to the angular velocity of rotation while
the gravielectric field corresponds to the temperature gradient.
In this paper we will estimate the vortical susceptibility of the low density QCD matter using the anomalous
triangle $<VAV_{g}>$ where $V_g$ is the fermion current interacting with an external graviphoton field.

The non-vanishing vortical susceptibility of matter generates an interesting anomalous 
term in the low-energy QCD effective action. This new anomalous term yields the
anomalous contribution to the angular momentum in the confined phase of QCD in the external 
rank-two field which 
may be called "superrotation".   
It is the vortical counterpart of the anomalous term in an external electromagnetic field which 
generates  the supercurrent in QCD \cite{gkh}.
Recall that  supercurrents in the gapped  phases are   specific
features of the superconducting and superfluid states of matter. They
are proportional to the gradient of the phase of the condensate and
to the magnitude of the condensate.
The chiral condensate breaks the global symmetry of the QCD Lagrangian,  
hence the situation is more
close to the superfluid case when the global $U(1)$ symmetry is broken. The possible superrotation in an external selfdual rank two  field is of the same nature.

The paper is organized as follows. In Section 2 we recall the properties
of magnetic susceptibility and introduce its vortical counterpart. 
It is evaluated via the mixed anomaly and the pion dominance in the axial channel.
In Section 3 we discuss the possibilities for related effects in more
general gravitational backgrounds and in the chirally imbalanced matter. 
The application of our results to $\Lambda$ polarization observed recently in heavy ion collisions is described  in  Section 4.

\section{ Susceptibilities of  the chiral condensate in the external fields}
\subsection{ Magnetic susceptibility}

Consider the hadronic phase of QCD where the chiral symmetry is broken by the
condensate $<\bar{\Psi}\Psi>$. The linear response of the 
chiral condensate to an external magnetic field results in the relation
\beq
<0|\bar{\Psi}_f\sigma_{\mu\nu}\Psi_f|0> = e_f\chi_f <\bar{\Psi}\Psi>F_{\mu\nu}
\label{sus}
\eeq
where $\chi_f$ and $e_f$ are the susceptibility and the charge  of the corresponding quark flavor $f$.
Its value can be derived from the anomalous $<VVA>$ triangle, $\chi= - N_c/(4\pi^2 f_\pi^2)$ 
\cite{vainshtein}. 

If we introduce a physical selfdual rank-2 field $ B_{\mu\nu}$, the non-vanishing magnetic susceptibility induces the following anomalous term
in the chiral Lagrangian  \cite{gkkv}
\beq
L_{anom}= \chi <\bar{\Psi}\Psi>F_{\mu\nu}B_{\mu\nu}\ {\rm Tr} \left[ {\rm B Q} (U+U^{-1}) \right] ,
\label{anom}
\eeq
where $B_{\mu\nu}$ is the source for the fermion
tensor current  $\bar{\Psi}\sigma_{\mu\nu}\Psi$ , B is the flavor matrix and Q is charge matrix. This term can be thought of as the modification 
of the quark mass in the external field. In four dimensions
\beq
\sigma_{\mu\nu}= \frac{1}{2}i \epsilon_{\mu\nu\alpha\beta} \sigma_{\alpha\beta}\gamma_5
\eeq
therefore the external field has to be selfdual, $ B_{\mu\nu}=\tilde{B}_{\mu\nu}$. 

It has been argued in \cite{gkh} that this anomalous term induces a kind of a supercurrent 
proportional to the condensate
\beq
J_{anom,\nu}=\chi <\bar{\Psi}\Psi>\partial_{\mu}B_{\mu\nu}
\eeq
which due to selfduality can be also written in terms of the  dual field $\tilde{B}$. We can trade the $B_{\mu\nu}$ field 
for a scalar field $\Phi$;   the anomalous  current then acquires the form identical to the superfluid current
\beq
J_{superfluid,\mu}= n_s (\partial_{\mu}\Phi - A_{\mu})
\eeq
where $n_s$ is the density of the superfluid component proportional
to the condensate and $\Phi$ is the condensate phase. In our case the field $\Phi$
is the scalar meson while $A_{\mu}$ is the composite  current.

\subsection{Vortical susceptibility of the quark condensate}

There is a well-known analogy between an external magnetic field and
the rotating frame manifested by the substitution $e_f \vec{B} \leftrightarrow \mu_f \vec{\Omega}$ that is based upon Larmor's theorem.  It is thus natural 
to introduce a response of the quark condensate in QCD to rotation, parametrized as 
\beq\label{vort_s}
<0|\bar{\Psi}_f\sigma_{\mu\nu}\Psi_f|0>= \chi_{g,f} <\bar{\Psi}_f\Psi_f>G_{\mu\nu} ;
\eeq
here we  describe rotation as the curvature of an external ``graviphoton" field  $G_{\mu\nu}$
and denote the corresponding susceptibility as $\chi_{g,f}$.

Let us recall that the graviphoton field
is introduced as the specific fluctuation of the background metric:
\beq
ds^2=(1+ 2\phi_g)dt^2 -(1-2\phi_g)d^2\vec{x} + 2\vec{A_g}d\vec{x}dt .
\eeq
The gravimagnetic field corresponds to the angular velocity of rotation at small velocity  
\beq
\vec{B}_g\propto \vec{\Omega}  ;
\eeq
however at large velocities the relation between the gravimagnetic field and 
the angular velocity is more complicated.

To evaluate the vortical susceptibility let us first 
recall the derivation of magnetic susceptibility of the quark condensate in \cite{vainshtein}.
It is convenient to consider the $<V(0)V(Q)A(-Q)>$ three-point function assuming
that one vector vertex corresponds to the 
external magnetic field. This three-point function can be treated
in two ways. First, assuming that $Q^2$ is large we can consider the OPE for the
product of  vector and axial currents; the first subleading term is 
$m\bar{\Psi}\sigma_{\mu\nu}\Psi / Q^4$, where m is the quark mass. 
On the other hand we can assume the pion dominance in the axial channel and make  
use of the anomalous $\pi_0 F\tilde{F}$ vertex. Expanding the propagator 
in the pion mass we get the subleading $m_{\pi}^2 / Q^4$ term that has to be compared 
to the $Q^{-4}$ term in the OPE. The comparison of the two terms and the use of the GMOR 
relation and (\ref{sus}) then immediately yields the expression for the magnetic
susceptibility $\chi = - N_c / (4 \pi^2f_{\pi}^2)$ found in \cite{vainshtein}.

Let us follow the same strategy in the gravimagnetic case. First let us recall 
that the graviphoton that is a component of the metric interacts with the
energy-momentum tensor. The gravimagnetic component of the connection $A_g$ is coupled
to the momentum of the fermion field via the term 
\beq
A_{g,i}\bar{\Psi}(\gamma_i D_0 +\gamma_0 D_i)\Psi
\eeq
At finite density $\mu$, there is a component of the momentum that does not involve 
the derivatives:
\beq
\mu A_{g,i}\bar{\Psi}\gamma_i\Psi
\eeq
this expression is known from the derivation of the CVE \cite{cve}. Therefore the weak  gravimagnetic field
interacts with the vector current similarly to the electromagnetic current, with the additional factor $\mu$ in the vertex.

Let us now compare two ways of
evaluation of the correlator: via  OPE for $A(Q)T_{0j}(0)V(-Q)$ and via 
the pion pole dominance in the axial channel, similarly to \cite{vainshtein}. Since 
the momentum in $T_{0j}$ vertex vanishes, the operator in the case if finite density QCD  reduces to the vector current 
multiplied the chemical potential $\mu$; therefore  we get the correlator that is similar to the
magnetic case. From the OPE side we get
the same operator $m\bar{\Psi}\sigma_{\mu\nu}\Psi / Q^4$
as the first subleading term -- therefore the OPE involves precisely the matrix element
we are looking for.

From the pion pole side we will now use the 
anomalous pion-gravity $\pi_0 V B_g$ vertex and expand the pion propagator 
in pion mass once again. The specific vertex which  involves the Goldstone particle,
electromagnetic field and the graviphoton field exists only in dense matter \cite{zhit1,zhit2} and
reads as 
\beq
\delta L = -N_C\sum_f \frac{e_f \mu_f}{4\pi^2} \epsilon^{\mu\nu \alpha \beta} \partial_{\mu}\pi A_{\alpha} \partial_{\nu}A_{g,\beta}
\eeq
This vertex allows us to make the comparison with the OPE calculation.
The leading term corresponds to the pion pole, and comparison of the two $Q^{-4}$ terms yields 
\beq\label{vort_v}
\chi_{g,f}=  \frac{N_c\mu_f}{\pi^2 f_{\pi}^2}
\eeq
where we have introduced the susceptibility for each flavor.
Of course, as in the magnetic case, the assumption of pion dominance brings in an uncertainty due to the omission of excited resonances; however we believe that this result provides a reasonable estimate.

\section{Anomalous term in the low-energy action and angular momentum}

What are the possible observable manifestations of the vortical susceptibility of the condensate?
To answer this question let us first note that the non-vanishing vortical susceptibility of the chiral condensate implies the following anomalous
term in the low-energy chiral Lagrangian:
\beq
L_{anom}= \chi_g <\bar{\Psi}\Psi>  G_{\mu\nu}B_{\mu\nu}TrB(U+U^{-1}) ,
\eeq
where B is the flavor matrix (the charge matrix does not enter this expression).
In the gravimagnetic case when $G_{ij}= 2  \epsilon_{ijk} \Omega_k$  instead of the anomalous current we get the anomalous
angular momentum induced by the external rank-two B field
\beq
J_i\propto \chi_g <\bar{\Psi}\Psi> \epsilon_{ijk}B_{jk} \ .
\eeq
As noted before, the rank-two B field is selfdual and can be traded for the scalar and the composite  current.
 
\vskip0.3cm

So far we have found the vortical susceptibility 
of the quark condensate which is proportional to the vector
chemical potential. One may also consider the case of the 
chirally imbalanced matter with chiral chemical potential $\mu_5\neq 0$. In this case due to the parity-odd nature 
of the chemical potential several new effects emerge. 
First,  one can introduce the  susceptibility in  the external gravielectric field that describes  
the temperature gradient:
\beq
<0|\bar{\Psi}\sigma_{\mu\nu}\Psi|0>= \tilde{\chi}_{g} <\bar{\Psi}\Psi> \tilde{G}_{\mu\nu} ,
\eeq
where the $\tilde{\chi}_{g}\propto \mu_5$ and the gravielectric field $E_g \propto \nabla \log T$.

In addition,  in chirally imbalanced matter we  
expect the ``anomalous boost" induced by the external rank-two field and given by
the derivative
of the Lagrangian with respect to the gravieletric field $G_{0j}$:
\beq
J_{j}= \frac{\delta L}{\delta G_{0j}} .
\eeq
This vector is dual to the angular momentum which is the axial vector. It is proportional
to the dual vortical susceptibility and the corresponding 
component of the rank 2 external field: 
\beq
J_{j} \propto \tilde{\chi}_g <\bar{\Psi}\Psi> B_{0j} .
\eeq

The chiral chemical potential can be generated by topological QCD fluctuations, or by a  
time-dependent axion field during the creation and decay of the extended axionic defects like
the axion string and axion domain wall.

\section{Vorticity-induced polarization throughout the hadronization: $\Lambda$ production in heavy ion collisions}

The vortical susceptibility of the quark condensate discussed above has a simple physical interpretation: vorticity aligns the spins of quarks and antiquarks along the axis of rotation. While the alignment of quark spins by rotation in a chirally symmetric phase has been extensively discussed in the literature \cite{Kharzeev:2004ey, Liang:2004ph, Liang:2004xn, zhit3, Betz:2007kg, Gao:2007bc, Becattini:2013vja, Jiang:2016wvv} (see \cite{Kharzeev:2015znc} for a recent review), {\it a priori} it is not clear whether any of this alignment would survive the hadronization. Indeed, hadronization is accompanied by the spontaneous breaking of chiral symmetry, and could thus easily wash out any spin polarization present in the chirally symmetric phase. Our results in this paper however indicate that the polarization of quarks and antiquarks induced by vorticity should survive the hadronization transition. This effect should manifest itself in spin polarization of the observed hadrons. 
\vskip0.3cm

Recently, the STAR Collaboration at RHIC has presented the preliminary data indicating that both $\Lambda$ and ${\bar \Lambda}$ hyperons in AuAu collisions at $\sqrt{s} = 62$ GeV and below appear polarized relative to the reaction plane, with $P_\Lambda \simeq  P_{\bar \Lambda} \simeq 0.02 - 0.03$ \cite{STAR}. We believe that this observation is remarkable for the following reasons: 
i) it presents a clear evidence for the presence of vorticity in the quark-gluon fluid; ii) it shows that vorticity is still present at the  hadronization transition, suggesting a small value of shear viscosity; iii) it demonstrates that the strange quark and antiquark polarization survives the hadronization. 

Let us now present the estimate of $\Lambda$ and ${\bar \Lambda}$ polarization basing on our results. Assume that the vorticity is aligned along the $z$ axis; the average spin polarization of quarks and antiquarks from (\ref{vort_s}) and (\ref{vort_v}) is
\beq\label{fin_res}
\langle \sigma_z \rangle = \frac{\langle {\bar \Psi} \sigma_{xy}\gamma_5 \Psi \rangle}{\langle {\bar \Psi} \Psi \rangle} = \frac{ N_c}{\pi^2 f_\pi^2}\ \mu\ \Omega_z ,
\eeq
where $\mu$ is the baryon chemical potential and $\Omega$ is the vorticity. 

For a numerical estimate, let us use the baryon chemical potential $\mu \simeq 80$ MeV as extracted from hadron abundances in AuAu collisions at $\sqrt{s} = 62$ GeV \cite{Braun-Munzinger:2015hba}. Since the polarization of $\Lambda$ hyperons results mostly from the polarization of strange quarks, we have to use the chemical potential of strange quarks that is $\mu_s \simeq \mu/3$.  The value of vorticity on the freeze-out hyper-surface at present is somewhat uncertain. The lower bound on this quantity can be obtained by making a (somewhat unrealistic) assumption that no vorticity is present in the initial conditions. One can then use viscous $(3+1)$ relativistic hydrodynamics with shear viscosity to entropy ratio $\eta/s = 0.1$ to obtain the lower bound of $\Omega_z \simeq 5 \cdot 10^{-3}\ {\rm fm}^{-1}$ \cite{Becattini:2015ska}. 
On the other hand, the evaluation of vorticity in the initial state \cite{Deng:2016gyh} from HIJING model yields for initial vorticity a large value $\Omega^0_z \simeq 0.1$. Hydrodynamical evolution at small shear viscosity is then expected to approximately  preserve vorticity, yielding the value of $\Omega_z \simeq 0.1\ {\rm fm}^{-1}$ on the freeze-out hypersurface \cite{inprog}. 

 With these values ($\mu_s \simeq 27$ MeV, $\Omega_z \simeq 0.1\ {\rm fm}^{-1}$), $N_c = 3$ and the pion decay constant $f_\pi = 92$ MeV we get from (\ref{fin_res}) the strange quark polarization 
\beq
P_s \equiv  \langle \sigma_z \rangle \simeq 0.02 .
\eeq
Of course, the polarization of $\Lambda$ and ${\bar \Lambda}$ will be suppressed by the dilution factor $D$ relative to the polarization of strange quarks and antiquarks:
\beq
P_\Lambda = P_{\bar \Lambda} = D\ P_s .
\eeq
The data on production of polarized $\Lambda$ and ${\bar \Lambda}$ hyperons in semi-inclusive Deep-Inelastic scattering allow to extract the dilution factor; in the kinematic region of valence quarks appropriate for our case, it is $D = 0.9 \pm 0.2$ \cite{Ellis:1995fc}.

We thus estimate
\beq
P_\Lambda = P_{\bar \Lambda} \simeq 0.02, 
\eeq
in reasonable agreement with STAR data \cite{STAR}. Of course, this agreement could be much worse if we assumed the  absence of vorticity in the initial state. We think however that this assumption is unrealistic; clearly, much more work has to be done to control the magnitude of vorticity.

\section{Conclusion}

In this paper we have introduced the vortical susceptibility of finite density matter in the chirally broken phase of QCD. 
It originates from the anomaly, and has been estimated here 
under the assumption of pion dominance in the axial channel.
The non-vanishing vortical susceptibility signals a kind of ``super-rotation" (anomalous contribution to the angular momentum) present in a background antisymmetric tensor or axion fields 
in the rotating frame. It would be interesting to investigate the vortical 
susceptibility on the lattice similarly to how it has been done previously for magnetic susceptibility. This quantity can also be evaluated in the holographic approach.

We discuss a macroscopic property of the QCD matter, and it would be interesting 
to establish the microscopic origin of gravimagnetization in the rotating frame.
Since the vortical susceptibility is proportional to the baryon density it is natural 
to assume that at microscopic level one deals with gravimagnetization of the degrees of freedom that carry baryon 
charge. This can be compared with the gravimagnetization
of the vacuum in the  $N=2$ SYM theory when microscopically the gravimagnetization of vacuum is dominated
by the instantons \cite{gorsky} and can be evaluated from Nekrasov partition function
at small parameters of the Omega-deformation. Recall that baryons are identified as the instantons
in the holographic QCD hence the SYM results could provide some guidance in understanding gravimagnetization
in QCD vacuum.

It would also be interesting to consider the vortical susceptibility beyond the 
linear response approximation. At large baryon chemical potential $\mu$ QCD is expected to be in the deconfined  and chirally symmetric phase, but one could consider the case of a moderate $\mu$ and large vorticity. In analogy with the magnetic field case one may expect a saturation of vortical  
susceptibility at some value of vorticity; at very large vorticity, the QCD matter has recently been argued \cite{Jiang:2016wvv} to undergo transition to the chirally symmetric phase. 
Another interesting direction is the application of these ideas in condensed matter physics, to systems possessing exciton condensates.  The exciton condensate is the 
analog of the chiral condensate, so one can address the response  of the exciton condensate to the rotation of the system.

\vskip0.3cm

We thank A. Abanov and N. Sopenko for discussions on vortical susceptibility, and F. Becattini, X. Guo, X.-G. Huang, J. Liao, and M. Lisa for valuable exchanges on the role of vorticity in heavy ion collisions.
A.G. thanks the
Simons Center for Geometry and Physics at Stony Brook University where this paper has been completed  for  hospitality and support 
during the program ``Geometry of the Quantum Hall State". D.K. acknowledges support of the Alexander von Humboldt foundation and Le Studium foundation, Loire Valley, France for support during the  ``Condensed matter physics meets relativistic quantum field theory" program. The work of D.K. was supported by the U.S. Department of Energy under Contracts No. DEFG-88ER40388
and DE-SC0012704.
The work of A.A., D.F.  was supported 
by grant RFBR----15-02-02092,
the work of A.G. by grant RFBR----15-02-02092   and grant of  
Russian Science Foundation 14-050-00150 for IITP.


\begin{thebibliography}{99}
\bibitem{zhit1}
 D.~T.~Son and A.~R.~Zhitnitsky,
  ``Quantum anomalies in dense matter,''
  Phys.\ Rev.\ D {\bf 70}, 074018 (2004)
  doi:10.1103/PhysRevD.70.074018
	\bibitem{zhit2} 
  M.~A.~Metlitski and A.~R.~Zhitnitsky,
  ``Anomalous axion interactions and topological currents in dense matter,''
  Phys.\ Rev.\ D {\bf 72}, 045011 (2005)
  doi:10.1103/PhysRevD.72.045011
  [hep-ph/0505072].
\bibitem{zhit3} 
  D.~Kharzeev and A.~Zhitnitsky,
  ``Charge separation induced by P-odd bubbles in QCD matter,''
  Nucl.\ Phys.\ A {\bf 797}, 67 (2007)
  doi:10.1016/j.nuclphysa.2007.10.001
  [arXiv:0706.1026 [hep-ph]].

	\bibitem{erdmenger} 
  J.~Erdmenger, M.~Haack, M.~Kaminski and A.~Yarom,
  ``Fluid dynamics of R-charged black holes,''
  JHEP {\bf 0901}, 055 (2009)
  doi:10.1088/1126-6708/2009/01/055
  [arXiv:0809.2488 [hep-th]].
	\bibitem{cve}
	K.~Landsteiner, E.~Megias and F.~Pena-Benitez,
  ``Gravitational Anomaly and Transport,''
  Phys.\ Rev.\ Lett.\  {\bf 107}, 021601 (2011)
  doi:10.1103/PhysRevLett.107.021601
  [arXiv:1103.5006 [hep-ph]].\\
K.~Landsteiner, E.~Megias, L.~Melgar and F.~Pena-Benitez,
  ``Holographic Gravitational Anomaly and Chiral Vortical Effect,''
  JHEP {\bf 1109}, 121 (2011)
  doi:10.1007/JHEP09(2011)121
  [arXiv:1107.0368 [hep-th]].
  
  \bibitem{Kharzeev:2010gr} 
  D.~E.~Kharzeev and D.~T.~Son,
  ``Testing the chiral magnetic and chiral vortical effects in heavy ion collisions,''
  Phys.\ Rev.\ Lett.\  {\bf 106}, 062301 (2011)
  doi:10.1103/PhysRevLett.106.062301
  [arXiv:1010.0038 [hep-ph]].
\bibitem{ioffe} 
  B.~L.~Ioffe and A.~V.~Smilga,
  ``Nucleon Magnetic Moments and Magnetic Properties of Vacuum in QCD,''
  Nucl.\ Phys.\ B {\bf 232}, 109 (1984).
  doi:10.1016/0550-3213(84)90364-X
	\bibitem{vainshtein} 
  A.~Vainshtein,
  ``Perturbative and nonperturbative renormalization of anomalous quark triangles,''
  Phys.\ Lett.\ B {\bf 569}, 187 (2003)
  doi:10.1016/j.physletb.2003.07.038
  [hep-ph/0212231].
\bibitem{gk} 
  A.~Gorsky and A.~Krikun,
  ``Magnetic susceptibility of the quark condensate via holography,''
  Phys.\ Rev.\ D {\bf 79}, 086015 (2009)
  doi:10.1103/PhysRevD.79.086015
  [arXiv:0902.1832 [hep-ph]].
	\bibitem{harvey} 
  S.~K.~Domokos, J.~A.~Harvey and A.~B.~Royston,
  ``Successes and Failures of a More Comprehensive Hard Wall AdS/QCD,''
  JHEP {\bf 1304}, 104 (2013)
  doi:10.1007/JHEP04(2013)104
  [arXiv:1210.6351 [hep-th]].
\bibitem{gkkv} 
  A.~Gorsky, P.~N.~Kopnin, A.~Krikun and A.~Vainshtein,
  ``More on the Tensor Response of the QCD Vacuum to an External Magnetic Field,''
  Phys.\ Rev.\ D {\bf 85}, 086006 (2012)
  doi:10.1103/PhysRevD.85.086006
  [arXiv:1201.2039 [hep-ph]].
\bibitem{buividovich}
P.~V.~Buividovich, M.~N.~Chernodub, E.~V.~Luschevskaya and M.~I.~Polikarpov,
  ``Chiral magnetization of non-Abelian vacuum: A Lattice study,''
  Nucl.\ Phys.\ B {\bf 826}, 313 (2010)
  doi:10.1016/j.nuclphysb.2009.10.008
  [arXiv:0906.0488 [hep-lat]].
	\bibitem{bali} 
  G.~S.~Bali, F.~Bruckmann, M.~Constantinou, M.~Costa, G.~Endrodi, S.~D.~Katz, H.~Panagopoulos and A.~Schafer,
  ``Magnetic susceptibility of QCD at zero and at finite temperature from the lattice,''
  Phys.\ Rev.\ D {\bf 86}, 094512 (2012)
  doi:10.1103/PhysRevD.86.094512
  [arXiv:1209.6015 [hep-lat]].
\bibitem{braun}
V.~M.~Braun, S.~Gottwald, D.~Y.~Ivanov, A.~Schafer and L.~Szymanowski,
  ``Exclusive photoproduction of hard dijets and magnetic susceptibility of QCD vacuum,''
  Phys.\ Rev.\ Lett.\  {\bf 89}, 172001 (2002)
  doi:10.1103/PhysRevLett.89.172001
  [hep-ph/0206305].
\bibitem{rohrwild} 
  J.~Rohrwild,
  ``Determination of the magnetic susceptibility of the quark condensate using radiative heavy meson decays,''
  JHEP {\bf 0709}, 073 (2007)
  doi:10.1088/1126-6708/2007/09/073
  [arXiv:0708.1405 [hep-ph]].
	
	\bibitem{gkh}
	A. Gorsky and D. Kharzeev, 
	"`On the Supercurrent in QCD Vacuum"', to appear
	
	\bibitem{gorsky}
 A.~Gorsky,
  ``Angular Momentum and Gravimagnetization of the ${\cal N}=2$ SYM vacuum,''
  Theor.\ Math.\ Phys.\  {\bf 171}, 616 (2012)
  doi:10.1007/s11232-012-0059-9
  [arXiv:1102.1841 [hep-th]].

\bibitem{Kharzeev:2004ey} 
  D.~Kharzeev,
  Phys.\ Lett.\ B {\bf 633}, 260 (2006)
  doi:10.1016/j.physletb.2005.11.075
  [hep-ph/0406125].

\bibitem{Liang:2004ph} 
  Z.~T.~Liang and X.~N.~Wang,
  Phys.\ Rev.\ Lett.\  {\bf 94}, 102301 (2005)
  Erratum: [Phys.\ Rev.\ Lett.\  {\bf 96}, 039901 (2006)]
  doi:10.1103/PhysRevLett.94.102301, 10.1103/PhysRevLett.96.039901
  [nucl-th/0410079].
  
\bibitem{Liang:2004xn} 
  Z.~T.~Liang and X.~N.~Wang,
  Phys.\ Lett.\ B {\bf 629}, 20 (2005)
  doi:10.1016/j.physletb.2005.09.060
  [nucl-th/0411101].

\bibitem{Betz:2007kg} 
  B.~Betz, M.~Gyulassy and G.~Torrieri,
  Phys.\ Rev.\ C {\bf 76}, 044901 (2007)
  doi:10.1103/PhysRevC.76.044901
  [arXiv:0708.0035 [nucl-th]].

\bibitem{Gao:2007bc} 
  J.~H.~Gao, S.~W.~Chen, W.~t.~Deng, Z.~T.~Liang, Q.~Wang and X.~N.~Wang,
  Phys.\ Rev.\ C {\bf 77}, 044902 (2008)
  doi:10.1103/PhysRevC.77.044902
  [arXiv:0710.2943 [nucl-th]].

\bibitem{Becattini:2013vja} 
  F.~Becattini, L.~Csernai and D.~J.~Wang,
  Phys.\ Rev.\ C {\bf 88}, no. 3, 034905 (2013)
  Erratum: [Phys.\ Rev.\ C {\bf 93}, no. 6, 069901 (2016)]
  doi:10.1103/PhysRevC.93.069901, 10.1103/PhysRevC.88.034905
  [arXiv:1304.4427 [nucl-th]].

  
\bibitem{Jiang:2016wvv} 
  Y.~Jiang and J.~Liao,
  arXiv:1606.03808 [hep-ph].

\bibitem{Kharzeev:2015znc} 
  D.~E.~Kharzeev, J.~Liao, S.~A.~Voloshin and G.~Wang,
  Prog.\ Part.\ Nucl.\ Phys.\  {\bf 88}, 1 (2016)
  doi:10.1016/j.ppnp.2016.01.001
  [arXiv:1511.04050 [hep-ph]].
	
	\bibitem{STAR} M. Lisa et al [STAR Collaboration], Talk at QCD Workshop on Chirality, Vorticity and Magnetic field in Heavy Ion Collisions, UCLA, January 2016; \\
	H. Huang et al [STAR Collaboration], Talk at the BEST 2016 Workshop, Indiana University, May 2016. 
	
\bibitem{Braun-Munzinger:2015hba} 
  P.~Braun-Munzinger, V.~Koch, T.~Sch�fer and J.~Stachel,
  ``Properties of hot and dense matter from relativistic heavy ion collisions,''
  Phys.\ Rept.\  {\bf 621}, 76 (2016)
  doi:10.1016/j.physrep.2015.12.003
  [arXiv:1510.00442 [nucl-th]].
	
\bibitem{Becattini:2015ska} 
  F.~Becattini {\it et al.},
  ``A study of vorticity formation in high energy nuclear collisions,''
  Eur.\ Phys.\ J.\ C {\bf 75}, no. 9, 406 (2015)
  doi:10.1140/epjc/s10052-015-3624-1
  [arXiv:1501.04468 [nucl-th]].

\bibitem{Deng:2016gyh} 
  W.~T.~Deng and X.~G.~Huang,
  arXiv:1603.06117 [nucl-th].
  
  \bibitem{inprog}
  X. Guo, X.G. Huang, and D. Kharzeev, {\it work in progress}.
  
\bibitem{Ellis:1995fc} 
  J.~R.~Ellis, D.~Kharzeev and A.~Kotzinian,
  ``The Proton spin puzzle and lambda polarization in deep inelastic scattering,''
  Z.\ Phys.\ C {\bf 69}, 467 (1996)
  doi:10.1007/s002880050048
  [hep-ph/9506280].
\end{thebibliography}
\end{document}